\documentclass[10pt,aps,prl,raggedbottom,longbibliography,reprint,citeautoscript,letterpaper,superscriptaddress]{revtex4-2} % reprint
\usepackage[usenames,dvipsnames]{color}
\usepackage{graphicx,microtype}
\usepackage{booktabs} 
\usepackage{multirow} 

\usepackage[bookmarks=false,colorlinks]{hyperref}
\hypersetup{linkcolor=magenta,citecolor=MidnightBlue,filecolor=Plum,urlcolor=MidnightBlue}
\usepackage[all]{hypcap}



\begin{document}

\title{Quantum sensing of magnetic fields with molecular color centers}

\author{Kathleen R.\ Mullin}
\affiliation{Department\,of\,Materials\,Science\,and\,Engineering, Northwestern University, Evanston IL, 60208, USA}

\author{Daniel W.\ Laorenza}
\affiliation{Department of Chemistry, Massachusetts Institute of Technology, Cambridge, Massachusetts 02139, USA}

\author{Danna E.\ Freedman}
\affiliation{Department of Chemistry, Massachusetts Institute of Technology, Cambridge, Massachusetts 02139, USA}

\author{James M.\ Rondinelli}
\email{jrondinelli@northwestern.edu}
\affiliation{Department\,of\,Materials\,Science\,and\,Engineering, Northwestern University, Evanston IL, 60208, USA}

\begin{abstract}
Molecular color centers, such as $S=1$ Cr(\textit{o}-tolyl)$_{4}$, show promise as an adaptable platform for magnetic quantum sensing. Their intrinsically small size, i.e., 1-2 nm, enables them to sense fields at short distances and in various geometries. 
This feature, in conjunction with tunable optical read-out of spin information, offers the potential for molecular color centers to be a paradigm shifting materials class beyond diamond-NV centers by accessing a distance scale opaque to NVs. 
This capability could, for example, address ambiguity in the reported magnetic fields arising from two-dimensional magnets by allowing for a single sensing technique to be used over a wider range of distances. 
Yet, so far, these abilities have only been hypothesized with theoretical validation absent. 
We show through simulation that Cr(\textit{o}-tolyl)$_{4}$ can spatially resolve proximity-exchange versus direct magnetic field effects from monolayer CrI$_{3}$ by quantifying how these interactions impact the excited states of the molecule. 
At short distances, proximity exchange dominates through molecule-substrate interactions, but at further distances the molecule behaves as a typical magnetic sensor, with magnetostatic effects dominating changes to the energy of the excited state. 
Our models effectively demonstrate how a molecular color center could be used to measure the magnetic field of a 2D magnet and the role different distance-dependent interactions  contribute to the measured field. 
\end{abstract}

\date{\today}
\maketitle

%\section{Introduction}
Precise measurements of magnetic fields are an important tool in understanding the spin properties and spatial distribution of magnetic fields in low dimensional materials. 
Over the past decade, intensive technique development in magnetic field sensing showcased the importance of coupling spatial resolution with minute sensitivity \cite{kuImagingViscousFlow2020,voolImagingPhononmediatedHydrodynamic2021,marchioriNanoscaleMagneticField2022}. 
Where techniques, such as  fluxgate \cite{ripkaNewDirectionsFluxgate2000a} and hall effect sensors \cite{zhengSensitivityEnhancementGraphene2017a,liSpinOrbitTorque2021}, are limited to micrometer scale sensing resolution,  
recent advances in quantum sensing have focused on optically detected magnetic resonance (ODMR) with color centers in semiconductors, e.g., diamond-NV centers \cite{taylorHighsensitivityDiamondMagnetometer2008,RevModPhys.89.035002}. 
These defect-based color centers feature two key advantages---optical read-out and a high sensitivity to local magnetic fields. 
In ODMR, the ground state spin is initialized with optical excitation, probed with microwave pulses to coherently drive the magnetic sublevels of a triplet ground state via a singlet excited state, and then optical read-out is used to determine the populations of the spin sublevels \cite{taylorHighsensitivityDiamondMagnetometer2008}. 
These color centers allow for nanoscale resolution of magnetic fields  \cite{hongNanoscaleMagnetometryNV2013}, but do not allow for intimate sensing of fields near the analyte, because defect qubits, like diamond-NV centers, operating as magnetic field sensors are spatially limited; they are difficult to bring in close proximity to the analyte as the defect is embedded in a crystal \cite{radtkeNanoscaleSensingBased2019a}, e.g., the smallest analyte-qubit distance reported is 9\,nm \cite{doi:10.1073/pnas.2114186119}. 
This limitation is functionally related to the nature of these defects, many of which would not be thermodynamically stable in isolation; indeed, placing defects closer to the surface leads to their removal or lower coherence times. 
Molecules are zero-dimensional and functionally all surface.
They may feature lower coherence times, but they are tunable and solution processable. 
This cross platform portability, in particular, enables post-synthetic processing of molecular color centers in thin film geometries to probe  magnetic fields with distances ranging from angstroms to micrometers. 
Recent work has even highlighted this functionality with self-assembled monolayers of radical-based spins \cite{https://doi.org/10.1002/adma.202208998}.
The molecular color center  Cr(\textit{o}-tolyl)$_{4}$ has a paramagnetic triplet $S=1$ ground state ($^{3}$A symmetry \footnote{While the isolated molecule has $C_{1}$ symmetry, we use the symmetry labels of the idealized tetrahedral structure.}) owing to the Cr$^{4+}$ center and a zero-field splitting that allows for optical addressability analogous to color centers in solids \cite{OpticallyAddressableMolecular,laorenzaTunableCr4Molecular2021a,PhysRevB.95.035207,dilerCoherentControlHighfidelity2020}. 
Here, light is used to resonantly excite an electron between the $M_{s}=0,\pm 1$ sublevels via a singlet excited state \cite{taylorHighsensitivityDiamondMagnetometer2008,mazeNanoscaleMagneticSensing2008a}.
When exposed to even tiny magnetic fields, these electronic states undergo a Zeeman shift in energy ($E'$) proportional to the intensity of the magnetic field ($B$), the free electron $g$-factor ($g_{e}$), the Bohr magneton ($\mu_{B}$), and the Planck constant ($h$) as 
%shown in Eq.\ref{eq.zeeman_shift}. 
%\begin{equation} \label{eq.zeeman_shift}
$E^\prime = g_{e} \frac{\mu_{B}}{h}SB\,,
$
%\end{equation}   
where $E^\prime$ is the difference in energy  of the first singlet excited state ($^{1}$E symmetry)  in the presence of a finite magnetic field ($E_{B, ^{1}\mathrm{E}}$) from that at zero field ($E_{B=0, ^{1}\mathrm{E}}$):
\begin{equation} \label{eq.energy}
E^\prime = E_{B, ^{1}\mathrm{E}} - E_{B=0,^{1}\mathrm{E}}\,.
\end{equation}  
While precise sensing of magnetic fields is then obtained by experimentally measuring shifts in the energy levels $M_{s}=0,\pm 1$ sublevels \cite{rondinNanoscaleMagneticField2012}, we use shifts in the $^{1}$E excited state as a surrogate in our model 
\footnote{Accurate quantitative changes in zero field splitting values are difficult to obtain with single-particle orbitals from DFT \cite{doi:10.1063/1.4790167}. Post-Hartree–Fock methods (i.e., complete active space self-consistent field methods) give more accurate results, but are %not suitable 
not tractable for the sensor-analyte system explored. %periodic simulation cell needed to explore the effects of a substrate on the molecule's electronic structure. 
This necessitates a surrogate value for zero field splitting. Here we use the $^{1}E$ singlet excited state that is involved in the excitation to the magnetic sublevels as a proxy.}. 

\begin{figure}
\centering
     \includegraphics[width=0.98\columnwidth]{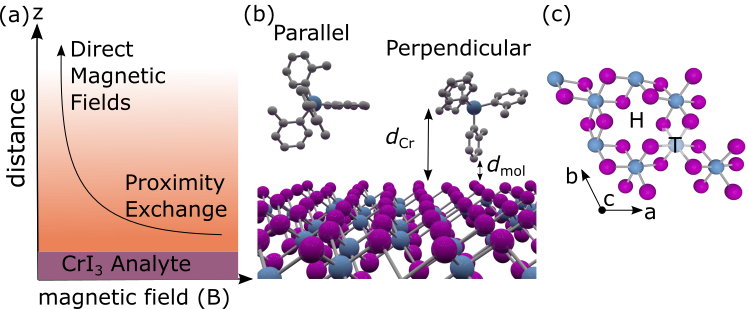}
    \caption{ (a) Schematic diagram showing the sensor-analyte geometry, whereby analyte dependent magnetic interactions can be probed by changing the distance between the senor and analyte. 
    Cr(\textit{o}-tolyl)$_{4}$ on CrI$_{3}$ can (b) 
    adopt two configurations in relation to the 2D magnet, i.e., with an \textit{o}-tolyl group parallel or perpendicular to the surface.
    (c) Hollow (H) and top (T) Cr adsorption sites on monolayer CrI$_{3}$.}
    \label{fig:geometry}
\end{figure}

\begin{figure*}
    \centering
    \includegraphics[width=1.75\columnwidth]{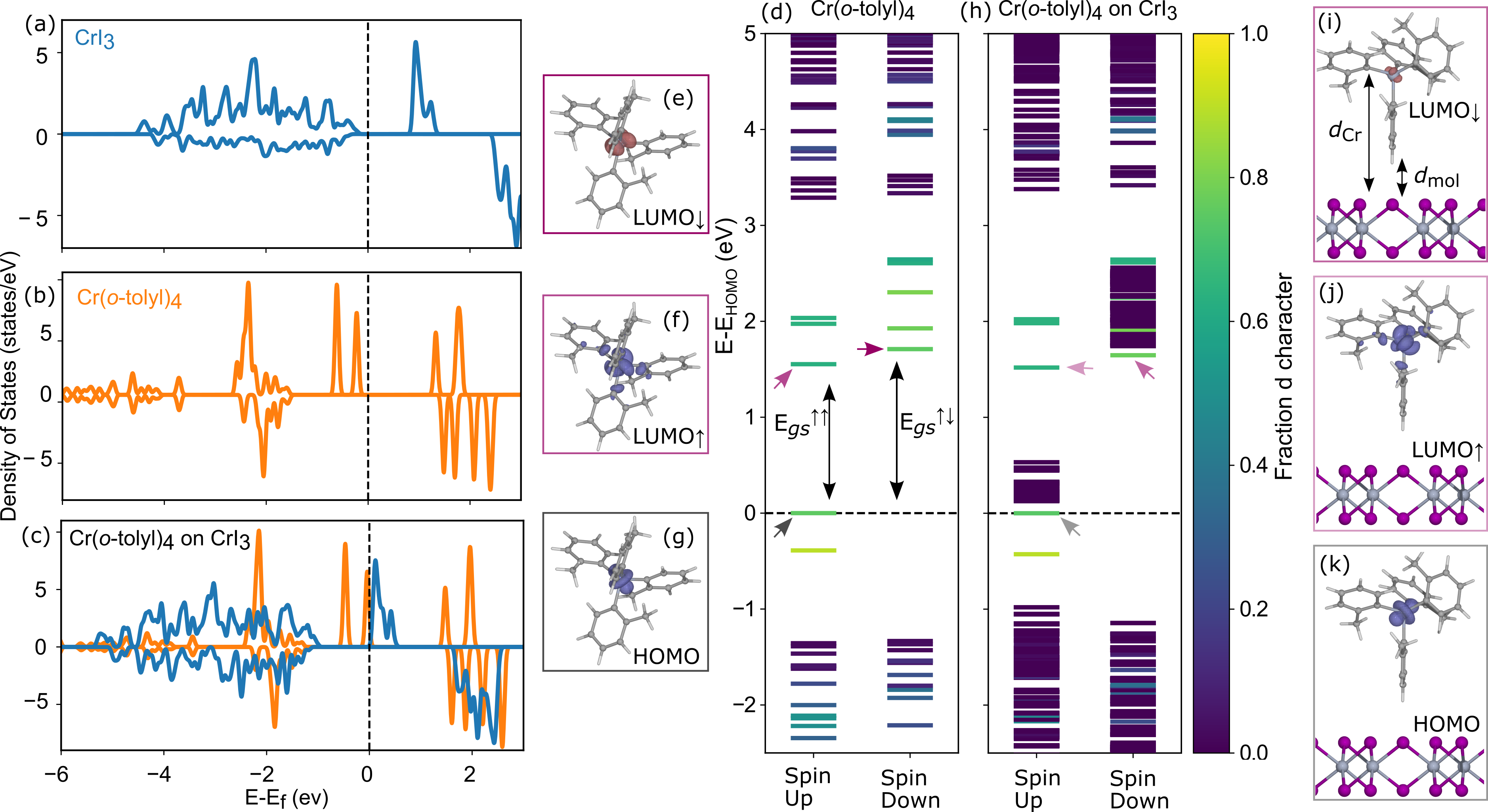}
    \caption{Electronic density of states (DOS) plots for (a) monolayer CrI$_3$, (b) Cr(\textit{o}-tolyl)$_{4}$ (b), and (c) Cr(\textit{o}-tolyl)$_{4}$ adsorbed on the CrI$_{3}$ surface (perpendicular geometry at the hollow site, see \autoref{fig:geometry}).  Quantitative molecular orbital diagrams highlighting the d-orbital character of (d) Cr(\textit{o}-tolyl)$_{4}$ and (h) Cr(\textit{o}-tolyl)$_{4}$ adsorbed on the CrI$_{3}$ surface. (e)-(gf) and (i)-(j) show visualizations of selected molecular orbitals for the isolated molecule and the molecule adsorbed on CrI$_{3}$, respectively, at $d_\mathrm{mol}=3$\,\AA\ ($d_\mathrm{Cr}=9$\,\AA). }
    \label{fig:DOS_MO_diagram}
\end{figure*}

Recently, many correlated 2D magnets such as ferromagnetic CrGeTe$_{3}$ and antiferromagnetic CrI$_{3}$, which exhibits layer dependent magnetic ordering below 45\,K, have been studied \cite{Gibertini_2019,Wang_2022}. 
To fully characterize these materials, high resolution sensors are needed. 
CrI$_{3}$ is an ideal analyte for quantum magnetic field sensing at the nanoscale, as prior studies have shown large discrepancies (five orders of magnitude) in magnetic fields with distance away from the surface of the monolayer (\autoref{fig:geometry}a). 
Zhong et al.\ used a CrI$_{3}$/WSe$_{2}$ heterostructure to calculate an effective magnetic field of 13\,T from the valley splitting in WSe$_{2}$ at 3.50\,\AA\ from CrI$_{3}$. 
Single spin microscopy with a diamond-NV center has  measured maximum magnetic fields of 0.31\,mT at a distance of 62\,nm from CrI$_{3}$ \cite{zhongVanWaalsEngineering, thielProbingMagnetism2D2019}. 
Additional studies on CrI$_{3}$-based heterostructures have shown that such large magnetic fields are due to proximity exchange 
%(direct overlap of the wavefunctions) 
rather than the typical magnetostatics  associated with stray magnetic fields \cite{zollnerGiantProximityExchange2020b,zollnerProximityExchangeEffects2019a}. 
This raises the question of how the 2D magnet distributes the magnetic flux, and this knowledge is needed to guide codesign of spin-based logic and memory devices \cite{Ahn_2020}. 

One approach to resolve the distance dependent magnetic field strengths is to layer thin films of Cr(\textit{o}-tolyl)$_{4}$, the sensor, of varying thicknesses on CrI$_3$, the analyte and perform thickness dependent ODMR measurements. 
This quantum sensing geometry in the single-molecule limit is shown in  \autoref{fig:geometry}a and it is the configuration we consider. 
(We discuss thin film effects later.)
We expect two important magnetic interactions, proximity exchange and direct magnetic fields, to occur between  Cr(\textit{o}-tolyl)$_4$ and the CrI$_{3}$ substrate.
Proximity exchange is a local effect and should dominate at short distances $d$ where electron wavefunctions directly overlap \cite{PhysRevB.106.035137}; in contrast, direct magnetic fields from diploar contributions should dominate at longer distances.  
Notably, it is very difficult to measure proximity exchange through conventional magnetic sensing techniques outside of NMR spectroscopy. 
Here, we use density functional theory (DFT) simulations  with magnetostatic calculations to show how the first excited state of Cr(\textit{o}-tolyl)$_{4}$ is affected by these interactions as a function of distance, demonstrating molecular color centers as a quantum magnetic sensing platform.  
%
%\section{Results/ Discussion}
We first performed electronic structure calculations on 
monolayer CrI$_{3}$ at the DFT-PBE level without spin-orbit coupling and in a ferromagnetic spin configuration %
\footnote{Ab-initio calculations were performed using the Vienna Ab-initio Simulation Package (VASP version 5.4.4) \cite{kresseInitioMolecularDynamics1993,kresseInitioMoleculardynamicsSimulation1994,kresseEfficiencyAbinitioTotal1996,kresseEfficientIterativeSchemes1996} using projector augmented wave pseudo potentials \cite{kresseUltrasoftPseudopotentialsProjector1999,blochlProjectorAugmentedwaveMethod1994} and the PBE exchange correlation functional \cite{perdewGeneralizedGradientApproximation1997,perdewGeneralizedGradientApproximation1996} with the following valence configurations I($5s^{2}5p^{5}$), Cr($3d^{5} 4s^{1}$), C($2s^{2}2p^{2}$), and H($1s^{1}$).   We used a 600\,eV energy cutoff for the planewave expansion and an energy convergence of $1\times10^{-7}$.
For $k$-point integrations, Gaussian smearing of 0.05\,eV was used.
}. 
\autoref{fig:DOS_MO_diagram}a shows there is a 1.1\,eV band gap and a large spin polarization, which leads to a local spin moment of 3.3\,$\mu_{B}$ on the Cr$^{3+}$ site consistent with prior monolayer studies \cite{wangSystematicStudyMonolayer2021}. 
Next, we simulated the electronic structure of 
Cr(\textit{o}-tolyl)$_{4}$ \footnote{For all systems involving the isolated molecule and the molecule on the substrate a single $k$-point at the $\Gamma$ point was used. For calculations on the 2D monolayer CrI$_{3}$ a $9\times9\times1$ $k$-point grid was used.
All components of the system were initially relaxed with a force convergence of 10$^{-3}$ meV\,\AA$^{-1}$ within in a simulation cell providing an spacing of 15 \AA \, between periodic images of the molecule in all directions or the substrate in the out of plane direction. 
Excited state (es) calculations were performed using the $\Delta$-SCF method \cite{doi:10.1063/1.1645787,PhysRevB.78.075441}. 
We calculate E$_{es}^{\uparrow\downarrow}$ by subtracting the total energy of a constrained occupancy calculation, where the electron in the the spin-up HOMO is promoted to the spin-down LUMO, from a standard ground state DFT calculation. For all calculations, including the constrained occupancy calculations, the relaxed ground state geometry is used.
Changes in E$_{es}^{\uparrow\downarrow}$ were then used to calculate the implied magnetic fields based on the Zeeman shift using %\autoref{eq.zeeman_shift} and 
\autoref{eq.energy}. 
}, 
which has the $^{3}$A ground state (gs) with both the d$_{z^{2}}$ and d$_{x^{2}-y^{2}}$ orbitals singly occupied (\autoref{fig:DOS_MO_diagram}b). 
The first $^{1}$E excited state involves a spin nonconserving transition to a doubly occupied d$_{x^{2}-y^{2}}$ orbital.  
\autoref{fig:DOS_MO_diagram}d shows that the highest occupied molecular orbital (HOMO), spin up lowest unoccupied molecular orbital (LUMO), and spin down LUMO have significant d character (\autoref{fig:DOS_MO_diagram}e-g). 
The energy difference between the HOMO and LUMO (E$_{gs}^{\uparrow\uparrow}$) is 1.52 eV, while the energy difference between the spin up HOMO and the spin down LUMO (E$_{gs}^{\uparrow\downarrow}$) is 1.64 eV.
We find in the gas-phase approximation that the excited state energy is 0.62\,eV, which %for the isolated molecule. This under predicts the energy compared to experimental results that
is underpredicted from recent experiments that approximate 
it as %the energy of the first excited state to be 
1.21\,eV \cite{laorenzaTunableCr4Molecular2021a}. 
This underestimation is expected based on our use of a semilocal density functional. 

Next, we created the combined analyte-sensor system. Cr(\textit{o}-tolyl)$_{4}$ is brought in proximity to the CrI$_{3}$ surface in a 
parallel and perpendicular configuration, as shown in  \autoref{fig:geometry}b. 
With these two configurations, we carried out adsorption site calculations with the metal center in  Cr(\textit{o}-tolyl)$_{4}$ in the hollow and top positions of CrI$_{3}$ (\autoref{fig:geometry}c). 
These calculations included van der Waals energy corrections implemented with the DFT-D3 method of Grimme \cite{grimmeConsistentAccurateInitio2010,grimmeEffectDampingFunction2011}.
We find that the lowest energy configuration 
is the perpendicular orientation of the molecule at  the hollow adsorption site at a distance of $d_\mathrm{mol}\sim3$\,\AA\ ($d_\mathrm{Cr}\sim9$\,\AA)  from the surface. 
These two distances, shown in \autoref{fig:geometry}b, give the distance between the surface atom in monolayer CrI$_{3}$ that is most extended in the out-of-plane  direction, $d_\mathrm{z}$, with the nearest atom or the Cr cation, respectively, in Cr(\textit{o}-tolyl)$_{4}$. 
The energies for other sites were similar with a difference in energy of 40 $\mu$eV between the lowest and highest energy configurations.
\autoref{fig:DOS_MO_diagram}c shows the overall  electronic structure of the constituent materials is largely unchanged in the equilibrium geometry for the combined system.
There is a rigid shift in states for CrI$_3$, owing to surface Fermi level pinning (charge neutrality level), that brings the  HOMO of Cr(\textit{o}-tolyl)$_{4}$ just below the conduction band of CrI$_3$. 
The semiconducting nature of monolayer CrI$_3$ persists with minor changes in bandwidth. %
Similarly, there is a minor increase in the width of the HOMO. 
\autoref{fig:DOS_MO_diagram}h presents the fraction d character for orbitals near the Fermi level for Cr(\textit{o}-tolyl)$_{4}$ adsorbed on CrI$_{3}$ in its equilibrium position.
\autoref{fig:DOS_MO_diagram}i-k show no significant change in character for the orbitals in the  Cr(\textit{o}-tolyl)$_{4}$ molecule. 
These findings cause us to deduce that Cr(\textit{o}-tolyl)$_{4}$ is physisorbed onto CrI$_3$. 
(The parallel geometry also leads to no noticeable changes in the 
electronic structure of either the molecular color sensor or analyte.)

\begin{figure}
    \centering
     \includegraphics[width=0.99\columnwidth]{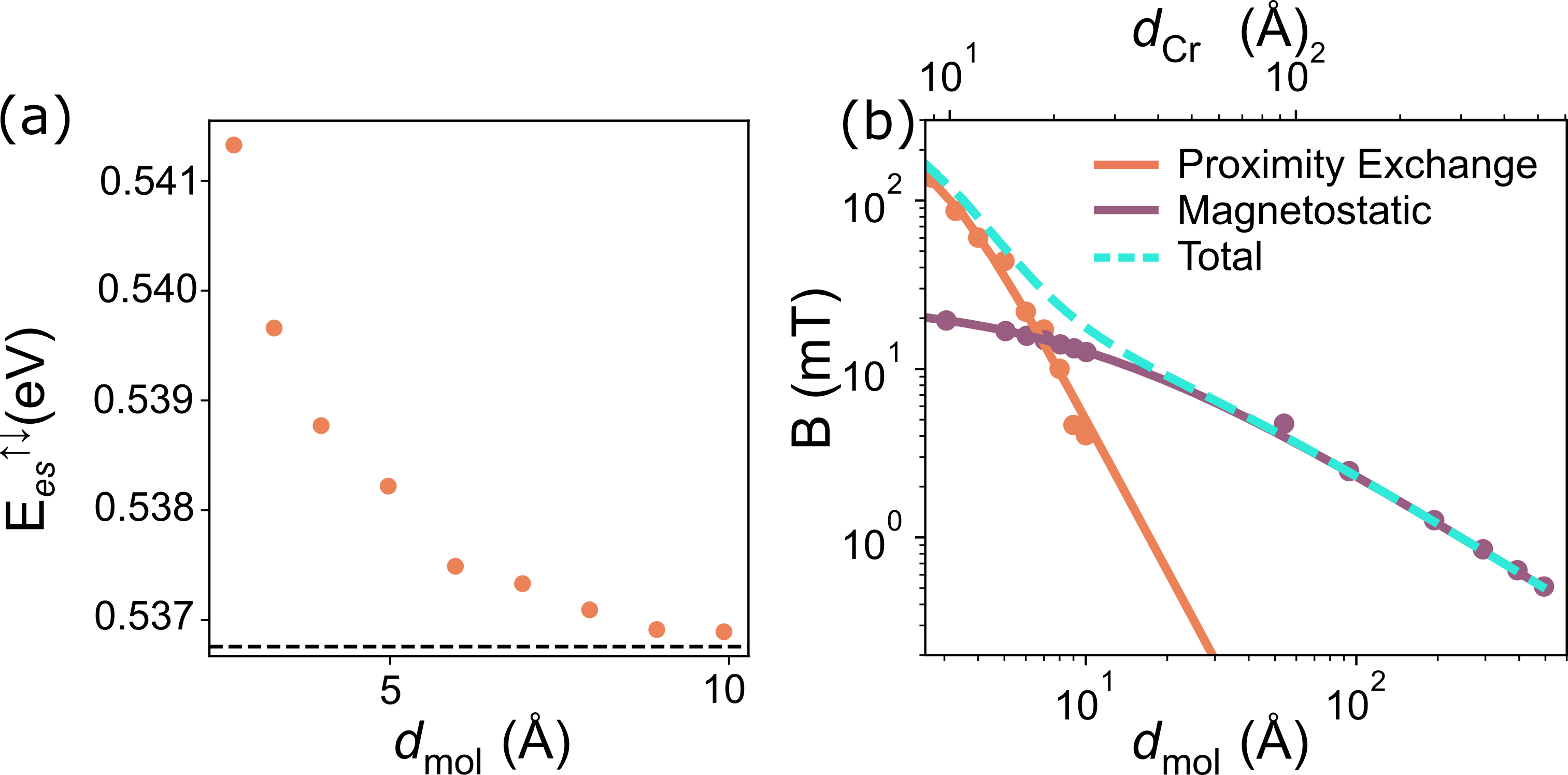}
    \caption{(a) Distance-dependent energies of the $^{1}$E excited state (E$_{es}^{\uparrow\downarrow}$) of the Cr(\textit{o}-tolyl)$_{4}$ in proximity to CrI$_{3}$. 
    The horizontal dotted line shows E$_{es}^{\uparrow\downarrow}$ for the isolated molecule. 
    (b) Total distance-dependent magnetic fields from CrI$_{3}$ and decompositions of the field into proximity exchange and magnetostatic contributions.
    Direct magnetic fields are quantized along the  easy axis of Cr(\textit{o}-tolyl)$_{4}$ (see text). 
    }
    \label{fig:model_comparison}
\end{figure}

In the aforementioned equilibrium geometry, we find that E$_{gs}^{\uparrow\uparrow}$ and E$_{gs}^{\uparrow\downarrow}$ increase slightly to 1.55 eV and 1.71 eV, respectively. E$_{es}^{\uparrow\downarrow}$ also decreases to 0.540 eV. 
We use these changes in the $^{3}$A ground state and $^{1}$E excited state energies of Cr(\textit{o}-tolyl)$_{4}$ to calculate the implied magnetic field sensed by the molecular color center through the Zeeman interaction. 
Here we obtain $B=94$\,mT at the equilibrium distance of 3 \AA.
We next examined the changes to these energies as a function of distance between between Cr(\textit{o}-tolyl)$_{4}$ and CrI$_3$ as means to map the magnetic field distribution transverse to the analyte. 
\autoref{fig:model_comparison}a shows that 
the energy of the first excited state decreases as Cr(\textit{o}-tolyl)$_{4}$ approaches CrI$_3$ over the range $2.7\le d_\mathrm{mol} \le 10$\,\AA.
This value falls off cubically as the distance increases, as shown in \autoref{fig:model_comparison}b, with the smallest computed field being 5.0\,mT at 10\,\AA.
As with the isolated molecule, these calculations do not account for relaxations in the excited state. 
Also, these %Data in \autoref{fig:mag_field_comparison}a for the proximity exchange result show the 
excited state energies of the molecule for the combined system are obtained relative to the structure of the relaxed molecule at the equilibrium distance without any geometric changes due to the change in distance from CrI$_{3}$; therefore, they represent the smallest changes expected as %differential 
atomic displacements would further increase the energy differences. 
We find the excited state energy of the isolated molecule changes from 0.621 eV in the fully relaxed geometry to 0.537 eV when the molecule's geometry is taken from the relaxed equilibrium combined structure.  
We use $E^\prime$ from the fully relaxed molecule to calculate an  effective magnetic field of $B=2.24$\,T at $d_\mathrm{mol}=2.7$\,\AA\ from the surface.
This may be compared to $B = 140$\,mT obtained at the 
same distance using $E_{B=0,^{1}E}$ from an unrelaxed geometry.
%
%shortest distance between Cr(\textit{o}-tolyl)$_{4}$ and CrI$_3$ of $d_\mathrm{mol}$ = 2.7\,\AA\ in the unrelaxed geometry.} 
%the implied magnetic field is calculated as B = 140\,mT.
%
This discrepancy is due to steric distortions to the inner shell of the Cr(\textit{o}-tolyl)$_{4}$ when adsorbed on the CrI$_{3}$ surface without charge transfer or bond formation.
We find a change in the coordination environment leading to an increase in the $\tau_{4}'$ geometry index, which quantifies the distortion of a 4 coordinate metal complex on a scale from 0 (square planar) to 1 (tetrahedral) \cite{Okuniewski_2015}, from 0.941 to 0.960 for the isolated molecule and the molecule on the surface, respectively. 
As this change is caused by the interaction of the molecule with the surface, we expect it to decrease with increasing distance between the molecule and the substrate.
Raman or infrared spectroscopy could be used to observe the structural distortion through shifts in the vibrational mode frequencies \cite{D1SC06130E}. 

\begin{figure}
    \centering
 \includegraphics[width=0.99\columnwidth]{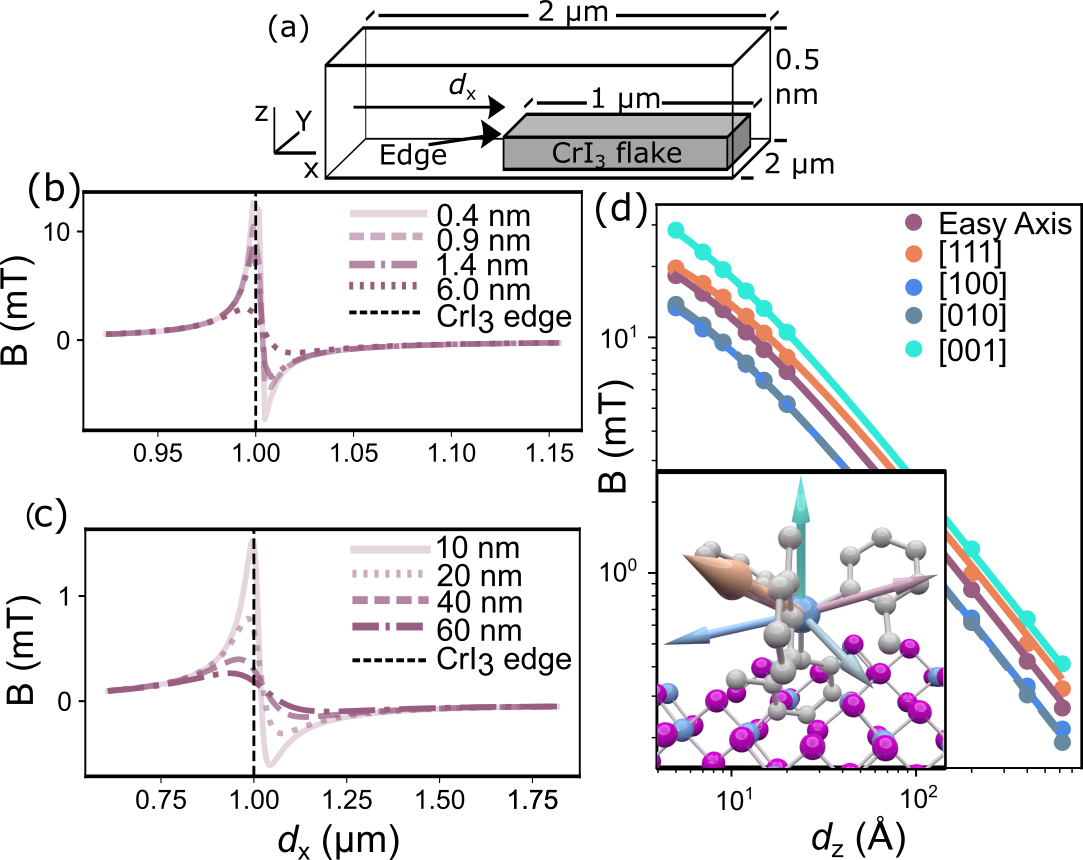}
    \caption{(a) Schematic of the simulation cell for the direct magnetic field models. (b,c) Computed magnetic fields from CrI$_{3}$ at variable $d_\mathrm{z}$ values. All results are taken along the line profile path shown in (a) and projected along the easy axis of the Cr(\textit{o}-tolyl)$_{4}$ molecule.  (d) Easy-axis orientation effects on the magnetostatic component of the sensed $B$ field. (inset) Vectors corresponding to orientations of the easy axis used. All vectors are projected onto the geometry of Cr(\textit{o}-tolyl)$_{4}$ molecule adsorbed on CrI$_{3}$.}
    \label{fig:direct_mag_field}
\end{figure}

We next computed the magnetic fields using a generalized model for the magnetostatic interactions from a parallelepipedic magnet \cite{ravaudMAGNETICFIELDPRODUCED2009a}, representing a coarse-grained model of  CrI$_3$ \footnote{Scripts for magnetostatic modeling are available at \url{https://github.com/MTD-group/Magnetostatic-Model-CrI3}}. 
Here, we modeled a freely suspended monolayer flake of CrI$_{3}$ with in-plane dimensions of $1\times1$\,$\mu$m within vacuum inside a $2 \times 2$\,$\mu$m area simulation cell in order to sample the field at the edge and above the flake (\autoref{fig:direct_mag_field}a).
The magnetic field was sampled every  7\,\AA\ over the plane of the simulation cell at vertical distances from the flake between  $5\le d_\mathrm{z} \le 600$ \AA. 
For input into the magnetostatic simulation, we modeled the magnetic polarization of CrI$_3$ using two methods:  
(1) a constant polarization calculated from an average of the DFT calculated spin density and
(2) a  full spin-density with each element in the spin density grid treated as a magnetic parallelepiped. 
We used the grid-based spin density approach (model 2) to calculate magnetic fields at distances less than 20\,\AA, while for greater distances we used the average calculated spin density (model 1). 
Both the grid based and the average spin density models 
quantitatively agree %lead to similar numerical result 
at distances ($>20$\, \AA) as shown in the Supporting Information \cite{Supp}.
The largest differences in magnetic fields from these two models occur for short distances, e.g., the difference in maximum field is 4.6 mT at 5\,\AA. This difference is much smaller than the 5 orders of magnitude discrepancy between previous experimental measurements.

\autoref{fig:direct_mag_field}b,c show that the magnetic field sharply increases in magnitude at the edge of the CrI$_3$ surface and then decreases to a small value over the top of the surface. 
The field then rapidly decreases to zero in the lateral direction beyond the CrI$_3$ boundary.
\autoref{fig:direct_mag_field}b,c further show that the peak magnetic field amplitude along the normal direction, $d_\mathrm{z}$, of monolayer CrI$_3$ increases and becomes more localized (decreasing peak width) with decreasing distance between Cr(\textit{o}-tolyl)$_{4}$ and the  analyte.
We find for $d_\mathrm{z}=60$\,nm, the magnetic field $B=0.28$\,mT.
This value closely matches the experimental maximum magnetic field of 0.31\,mT reported at the same distance \cite{thielProbingMagnetism2D2019}, demonstrating fidelity of the 
molecular color center is comparable to state-of-the-art diamond-NV centers. 
We compute the corresponding changes in energy for these magnetic fields from 
$E^\prime = g_{e} \frac{\mu_{B}}{h}SB$\,,
%\autoref{eq.zeeman_shift}, 
and find for the smallest fields $E^\prime=9.5$\,$\mu$eV, which are within the resolution of current ODMR measurements \cite{OpticallyAddressableMolecular,thielProbingMagnetism2D2019}.
%ODMR studies on solid and molecular color centers report energies down to 0.01 GHz (~1e-8 eV)
%%

%%
\autoref{fig:model_comparison}b shows the total magnetic field sensed as a function of distance decomposed into contributions from proximity exchange and the magnetostatics. 
Proximity exchange contributions are dominate up to 8 \AA\ from CrI$_{3}$. 
The short span over which  proximity exchange is the main contribution to the total magnetic field is due to the cubic distance dependent, $d^{3}$, decrease in proximity exchange compared to the $d^{-1}$ decrease for  direct magnetic fields.
While our analysis utilizes a single molecular color center, experimental setups would require molecular films.  
By comparing the distances where proximity exchange is the main contribution to the magnetic field to the inter-molecule spacing in the bulk molecular crystal ($\approx 9$\,\AA), we can extrapolate and determine the interaction a monolayer molecular film of Cr(\textit{o}-tolyl)$_{4}$ senses.
Here, proximity exchange would dominate the magnetic signal. 
In film thicknesses greater than these, the magnetostatic contributions would dominate the signal with proximity exchange having no measurable component.   
Experimental molecular films may also exhibit textures and grain boundaries.
Such  microstructure could elicit separate responses from distortions in the geometry of the molecule, which has a large effect on the excited state energy.
This could lead to increased variance in energy changes, adding complexity to interpretation of ODMR results.
Although the magnetic field from CrI$_{3}$ is a vector quantity, these sensors will yield a scalar value for the field that has been projected along the easy axis of the spin qubit.
The axis of quantization is important in both understanding the field magnitude for our single color center model and how molecular conformations in deposited films, further from the analyte, modify the alignment of the molecule's easy axis with respect to the direction of the magnetic field, thereby altering sensitivity. 
To  understand how these factors affect the fidelity of the field read-out, we next determined the spin quantization axis by performing noncollinear relativistic DFT calculations with spin-orbit coupling on  isolated Cr(\textit{o}-tolyl)$_{4}$. 
The energy of the easy axis aligned with the z axis, (-0.784, 0.500, 0.368), with respect to the combined molecule/substrate simulation cell was lower by 3.9\,$\mu$eV compared to the x axis and lower by 4.0\,$\mu$eV compared to the y axis (\autoref{fig:direct_mag_field}d, inset). 
\autoref{fig:direct_mag_field}d shows that the magnetic field decreases with $d_\mathrm{z}$ independent of the quantization axis, and these values are in good quantitative agreement with experiment.
There is, however, a strong dependence of the field strength on the axis of quantization, e.g., the largest different being 14 mT at 2.7\,\AA\ from the substrate.

Tailoring the easy axis of the sensor to the direction of the field would allow for maximum sensitivity. 
We now use this approach to quantize the 
simulated magnetic fields along the easy axis of a diamond-NV center and compare to the molecular color center. 
We find that they are within 20 $\mu$T at $d_\mathrm{z}=60$\,nm.
Comparing the magnetic field quantized along the easy axis from the Cr(\textit{o}-tolyl)$_{4}$ on CrI$_{3}$, to the field quantized along the [111] easy axis of a diamond-NV center, shows similar field strengths with the diamond NV center slightly larger. 
Neither the lowest energy orientation of Cr(\textit{o}-tolyl)$_{4}$ nor the known easy axis of the NV center are aligned with the axis having the largest magnetic field for CrI$_{3}$ (the z axis).
 The flexibility of molecular color centers both in the ligand and matrix may be useful variables to tune in maximizing the sensitivity of the sensor.

% 
%\jmr{This paragraph could be expanded to discuss magnetic fields and temperature scales. I think you need to discuss temperature scales relevant for magnetic materials where this sensing is relevant. The use case you picked is 2D magnets, so start there and summarize ordering temperatures for 2D magnets are these compatible where T1 times have been measured. Then you should expand to other quantum magnetic materials, like spin liquids, Kitaev materials \cite{Trebst_2022}, superconducting qubits, e.g., ask graham about his work, this could be used to sense magnetic vortices and resonators to look for inhomogeneites that lead to qubit decoherence and pair breaking. etc. Think a bit bigger about the implications of having a low-T quantum sensor. for metrology.}
%
Cr(\textit{o}-tolyl)$_{4}$ has a highly temperature dependent relaxation time varying from 500 to 3 $\mu$s between 5 and 40 K, respectively, which necessitates sensing at low temperatures \cite{laorenzaTunableCr4Molecular2021a,OpticallyAddressableMolecular}. 
Our use case for the sensor shown here, monolayer magnets, also have low ordering temperatures (45\,K for CrI$_{3}$ and 42\,K for CrGeTe$_{3}$) \cite{tiwariMagneticOrderCritical2021,Burch_2018,PhysRevB.105.245153}, making the molecular color center an ideal  sensing platform. 
Other potential use cases for these sensors include Kitaev materials, a SOC-entangled subclass of Mott insulators, such as RuCl$_{3}$, which exhibits unique zig-zag magnetic ordering below 7\,K \cite{Trebst_2022,PhysRevB.91.144420}. 
Additionally, magnetic defects, that may harm the performance of superconducting qubits via  decoherence channels, could be spatially mapped, as these magnetic defects are known to occur below 12\, K \cite{C4TC02222J,https://doi.org/10.48550/arxiv.2111.11684}. 
Sensing higher temperature phenomena will require increasing relaxation times, and there are substantial efforts focused on engineering ligand environments in molecular color centers to expand the useful temperature ranges to above liquid nitrogen  \cite{laorenzaTunableCr4Molecular2021a,FIELDING200692,doi:10.1021/acs.inorgchem.7b02616,D1SC06130E}.   

%\section{Conclusion}
In summary, we find that magnetic fields range from 94\,mT to 0.28\,mT %lead to 4\,meV changes in the first excited state 
over 3\,\AA\ to 60\,nm, respectively, using a single-sensing platform. 
We predict that at distances beyond 8\,\AA\ from the 2D magnet,  direct magnetic fields, i.e., the primary field probed by traditional magnetic field sensors, dominates the signal over the shorter range proximity exchange interaction. 
Our results show how molecular color centers can be used to sense magnetic fields with high fidelity over a range of distances, and how the platform forms a novel metrology to discern phenomena in low dimensional systems. 
The inherent flexibility of molecular color centers suggest alterations to the molecule could allow for optimizing the easy axis direction and the orientation of the molecule on the surface to allow for improved sensing.

%%%%%%%%
\begin{acknowledgments}
This work was supported by the U.S.\ Department of Energy, Office of Science, Basic Energy Sciences under award DE-SC0019356.
This work used resources at the National Energy Research Scientific Computing Center, a DOE Office of Science User Facility supported by the Office of Science of the U.S.\ Department of Energy under Contract No.\ DE-AC02-05CH11231.
\end{acknowledgments}

%%%%%%%%%%%%%%%%%%%%%%%%%%%%%%%%%%%%%%%%%%%%%%%%%%%%%%%%%%%%%%%%%%%%%%%%%%%%%%%%%%%%%%%%%%%%
\bibliography{references}

\end{document}